\documentstyle[prb,epsf,floats,aps,amsfonts,twocolumn,eqsecnum]{revtex}
\begin{document}
\draft
\wideabs{
\title{Connecting polymers to the quantum Hall plateau transition}
\author{Joel E. Moore}
\address{Bell Labs Lucent Technologies, 600 Mountain Avenue,
Murray Hill, NJ 07974}
\date{\today}
\maketitle
\begin{abstract}
A mapping is developed between the quantum Hall plateau transition and
two-dimensional self-interacting lattice polymers.  This mapping is
exact in the classical percolation limit of the plateau transition,
and diffusive behavior at the critical energy is shown to be related
to the critical exponents of a class of chiral polymers at the
$\theta$-point.  The exact critical exponents of the chiral polymer
model on the honeycomb lattice are found, verifying that this model is
in the same universality class as a previously solved model of
polymers on the Manhattan lattice.  The mapping is obtained by
averaging analytically over the local random potentials in a
previously studied lattice model for the classical plateau transition.
This average generates a weight on chiral polymers associated with the
classical localization length exponent $\nu = 4/3$.  We discuss the
differences between the classical and quantum transitions in the
context of polymer models and use numerical results on higher-moment
scaling laws at the quantum transition to constrain possible polymer
descriptions.  Some properties of the polymer models are verified by
transfer matrix and Monte Carlo studies.
\end{abstract}
}

\section{Introduction}

The quantum Hall plateau transition is of great interest because it
links Anderson localization and the quantum Hall effect (QHE), two of
the fundamental phenomena of condensed matter physics.  Noninteracting
electrons moving in two dimensions in a random potential form
energy eigenstates which do not extend to infinity but are
exponentially localized in finite regions of the plane.  The situation
changes drastically in a magnetic field: there are then
extended states at discrete critical energies $E_c$ separated by the
cyclotron energy $\hbar \omega_c$.  These extended states are remnants
of the Landau bands at zero disorder.  At other energies the electron
eigenstates are localized, and the localization length near a critical
energy scales according to
\begin{equation}
\xi(E) = \xi_0 \left({E_c \over E - E_c}\right)^\nu.
\label{nudef}
\end{equation}
An understanding of this behavior, which determines the passage from
one quantum Hall plateau to another, is essential to the explanation
of the integer QHE~\cite{laughlin}.  Despite progress in finding an
effective theory for this transition~\cite{pruisken,zirnbauer}, the
scaling law (\ref{nudef}) and other universal properties of the
transition have still not been obtained analytically.

The clearest picture for the scaling (\ref{nudef}) remains the
connection between hulls of percolation clusters and classical
electron trajectories in a strong magnetic field and random
potential~\cite{trugman}. Classical percolation was recently shown to
describe correctly some aspects of the {\it spin} quantum Hall
transition~\cite{gruzberg}, but the ordinary quantum Hall transition
is known from numerical studies~\cite{huckestein,chalker} not to lie
in the percolation universality class.  Considerable effort has been
devoted to how quantum effects modify the percolation picture, and
while there is now an understanding via numerics of the essential
ingredients required to model the transition~\cite{chalker}, analytic
progress on generalizing percolation has been quite
limited~\cite{dassarma}. This paper develops a mapping between the
plateau transition and the physics of self-interacting two-dimensional
lattice polymers.  Many statistical properties are known for such
polymers because they in turn can be mapped to simple magnetic systems
with known critical properties~\cite{degennes,nienhuis}.  The final
section discusses how quantum effects might be incorporated in a
polymer description.  However, numerically determined scaling laws for
higher moments in the quantum case, which result from strong
interference of quantum paths, seem to rule out a simple polymer
mapping.

The properties of a lattice version of the classical plateau
transition are mapped after disorder averaging onto polymers at the
$\theta$-point, which is a tricritical point separating the collapsed
and extended phases of polymers with an attractive short-ranged
self-interaction.  The polymer model for the classical transition
gives a useful complementary picture to the percolation description,
in which some facts, such as diffusion at the critical energy, are
more easily recovered.  A side benefit is that some new results on
polymers come out naturally from the mapping to the plateau
transition.

The polymer mapping provides an alternate connection between the
classical limit of the plateau transition and (classical)
percolation~\cite{trugman,evers,gurarie}, as ring polymers at the
$\theta$-point are equivalent to percolation
hulls~\cite{coniglio,duplantier}.  One result of this paper is that
directly mapping the plateau transition to polymers without the
intermediate step of percolation gives many more relations.  The
exponent $\nu_\theta$ which governs the typical size of a polymer
(for a polymer of $N$ units, $\langle R^2 \rangle \sim N^{2\nu_\theta}$,
where $\langle \rangle$ denote averages over the ensemble of
polymers), and $\mu$ and $\gamma$ which determine essentially the
number $\sim \mu^N N^{\gamma-1}$ of polymers of length $N$, can all
be connected to the plateau transition.

The localization length exponent in (\ref{nudef}) is $\nu =
\frac{4}{3}$ for classical percolation, while numerical
studies~\cite{huckestein,chalker} for the quantum case predict $\nu =
2.35 \pm 0.05$ for the lowest Landau level (LLL), consistent with
experiments~\cite{wei}.  In this paper $\nu$ will be studied via the
subdiffusive propagation of electrons in a magnetic field and quenched
random potential, which is now reviewed.  Recently it was shown by
Sinova, Meden, and Girvin~\cite{sinova} that the localization length
exponent $\nu$ appears in the energy-integrated correlation function
$\Pi(x,t) \equiv \langle \langle {\bar \rho}(0,0) {\bar \rho}(x,t)
\rangle\rangle$, where ${\bar \rho}$ is the LLL-projected electron
density operator and $\langle\langle\rangle\rangle$ denote disorder
averaging.  (The discussion in this paper is generally restricted to
the LLL, although $\nu$ is believed to be universal.)  The Fourier
transform was verified numerically to have the scaling form
\begin{equation}
\omega {\rm Im} \Pi(q,\omega) = \omega^{1 \over 2 \nu} f(q^2/\omega)
\label{scalform}
\end{equation}
in the limit $q,\omega \rightarrow 0$ with $q^2 \ll \omega$.
This scaling form can be understood as resulting from a simple
form for $\Pi(q,\omega)$ in this limit,
\begin{equation}
\Pi(q,\omega) \sim {1 \over \omega - i D(\omega) q^2},
\end{equation}
with the frequency-dependent diffusion constant $D(\omega) \propto
\omega^{1 \over 2\nu}$.

The result (\ref{scalform}) depends on the assumption that only at
isolated critical energies $E_c$ are there extended states.  It can be
understood from the following argument, which is somewhat different
from that in \cite{sinova}.  Electrons at energy $E$ with localization
length $\xi(E)$ move diffusively over short times but cross over to
localized behavior once $t \geq \xi(E)^2 / D_0$.  The diffusion
constant $D_0$ should have a finite limit as $E \rightarrow E_c$ since
the conductivity $\sigma_{xx}$ is finite at the transition, and can be
approximated by this limiting value in the scaling limit.  So for a
particle at the origin at $t=0$ (where it projects onto eigenstates of
different energies),
\begin{equation}
\langle x^2(t) \rangle \approx \int_{\xi(E) =
\sqrt{D_0 t}}^\infty dE\,\rho \xi^2(E)
+ \int_{E_c}^{\xi(E) = \sqrt{D_0 t}}dE\,\rho D_0 t.
\end{equation}
where $\rho$ is the density of states near $E_c$.  This results
in subdiffusive behavior:
\begin{equation}
\tilde D \equiv {d \over dt} \langle x^2(t) \rangle = \rho E_c D_0
\left(\xi_0 \over \sqrt{D_0} \right)^{1 \over \nu} t^{- {1 \over 2 \nu}},
\label{anomdiff}
\end{equation}
which corresponds to (\ref{scalform}) with $f(x) \propto x$ for $x \ll 1$.

Our starting point to obtain the anomalous diffusion (\ref{scalform})
is a single electron moving either classically
or quantum-mechanically in the $x$-$y$ plane in a random potential $V(x)$
and strong constant magnetic field $B {\bf \hat z}$.  The classical
coarse-grained equation of motion
\begin{equation}
B {\bf \dot x}_i = -\epsilon_{ij} \partial_j V(x)
\label{eom}
\end{equation}
can also be obtained~\cite{gurarie} by taking a certain limit of the
Liouvillian formalism.  The lowest-Landau-level projected electron
density operator in this limit becomes a classical distribution
function of particles moving according to (\ref{eom}).  Of course, the
equation of motion (\ref{eom}) can also be derived simply from
classical physics: a single electron moving in constant electric and
magnetic fields with $E < B$ has average velocity $\frac{E}{B}c$ along
the direction ${\bf E} \times {\bf B}.$ Since the direction of motion
is always perpendicular to ${\bf \nabla} V$, the particle
moves along a constant-energy contour of the potential.  The picture
underlying network models~\cite{chalker} of the transition is that
electron propagation is nearly classical except near a saddle point of
the potential, where quantum tunneling becomes significant.

The first part of this paper shows that a discrete-time lattice
version of (\ref{eom}), known to have the correct (percolative)
critical scaling for the classical limit, maps after disorder
averaging onto a model of two-dimensional interacting polymers on the
same lattice.  Although the lattice is useful to derive the mapping,
the critical properties related by the mapping are universal and hence
lattice-independent.  In the remainder of the introduction, we outline
the lattice model of (\ref{eom}) and some basic properties of
interacting polymers, then summarize the main results.

\begin{figure}
\epsfxsize=2.5truein
\centerline{\epsffile{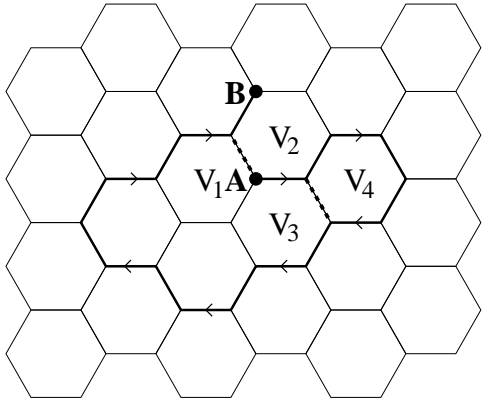}}
\caption{Sample self-avoiding walk from $A$ to $B$ of 20 steps,
with 19 neighboring hexagons.  The dotted edges are self-contacts:
the edge between $V_3$ and $V_4$ is an antiparallel self-contact,
while that between $V_1$ and $V_2$ is a parallel self-contact.
This path is not allowed classically since the walk passes
$V_2$ both on the right and on the left.}
\label{figone}
\end{figure}

In order to establish the connection between polymers and motion along
level surfaces, we use a lattice model due to Gurarie and
Zee~\cite{gurarie}.  The particle is taken to have constant velocity
along level surfaces: a nonzero mean velocity at criticality was found
numerically in \cite{evers,gurarie} for similar models, and fixing the
particle velocity does not alter the critical scaling.  Particles move
on the edges of the honeycomb lattice of Fig.~\ref{figone}, where each
hexagonal face has an associated random potential energy.  Except in
section II, the energy $E$ of a particle starting at vertex $A$ will
be taken to be the average of the three neighboring potentials $V_1,
V_2, V_3$, instead of an independent quantity as in~\cite{gurarie}.
The energy $E$ is constant along the particle trajectory.  The
particle's first step is chosen so that the potential to the left
is larger than the particle energy $E$, which is larger than the potential
to the right.  In successive steps, there is always a choice
between two directions aside from the direction by which the particle
entered, and only one of these choices will satisfy the condition that
the energy to the left (right) be greater (less) than $E$.  For each
realization of the random potentials and each starting point, there is
a unique locus of the particle after $N$ steps.  The connection to the
classical localization exponent $\nu=\frac{4}{3}$ is that the mean
square displacement after $N$ steps is found to show subdiffusive
behavior:
\begin{equation}
\langle\langle R^2(N) \rangle \rangle \sim N^{1 - \frac{1}{2 \nu}}
\label{polynu}
\end{equation}
in accord with (\ref{anomdiff}).  The value $1 - \frac{1}{2 \nu} \approx
0.62$ was found by numerical simulation~\cite{gurarie} of (\ref{eom}),
compared to the predicted value $\frac{5}{8} = 0.625.$

Now the connection between motion along level surfaces (\ref{eom}) and
percolation hulls is quickly reviewed.  Consider the level-surface
motion on the hexagonal lattice with white-noise potentials (i.e., the
disorder correlation length $\lambda_D$ is less than the lattice
spacing).  Then all potentials on the left of a trajectory are higher
than the trajectory energy $E$, which is higher than all potentials on
the right.  Now let all faces with energy higher than $E$ be
``colored'', while those with energy lower than $E$ remain uncolored.
The trajectory is then a hull separating colored faces from uncolored
ones, and the properties of such hulls are a standard problem in
percolation.  Although the choice of lattice affects such properties
as the density of colored faces at the critical point, critical
exponents are universal (independent of the lattice).  Similarly a
lattice model is used here to establish the mapping to a polymer
problem, but the universal polymer properties $\gamma$ and $\nu$
discussed below do not depend on this lattice.

The usual way to study this type of lattice model~\cite{evers,gurarie}
is by summing numerically over paths at fixed particle energy.  Here
the path will be held fixed for the integration over random
potentials: the goal is to assign a weight to each path according to
the fraction of the space of random potentials for which that path is
the particle trajectory.  For convenience, the potentials are assumed
to be uniformly distributed on $[-1,1]$, so the critical energy is
$E_0 = 0$.  After $N$ steps the particle has either moved along a
self-avoiding walk (SAW) of length $N$, or else has looped and begun
retracing previous steps.  There is a constraint of ``no parallel
self-contacts'' (the terminology is explained in section II) on
allowed SAWs resulting from the restriction that no hexagon can be
passed on both the left and right (Fig.~\ref{figone}).

The probability $P(a,b,N)$ to reach $b$ from $a$ after $N$ steps can
be written as a sum over an ensemble of paths including both
closed self-avoiding polygons (SAPs) as well as open SAWs with no
parallel self-contacts.
Writing $W^i$ for the weight of closed or open curve $i$,
the disorder-averaged probability to be at
$b$ after $N$ steps starting from $a$ is ($H$ is the number of different
hexagons visited by the SAW or SAP)
\begin{eqnarray}
P(a,b,N) &\propto& \sum_{{\scriptstyle{\rm SAPs\,}i{\rm \,through\,}
a {\rm\,and\,} b,\atop
\scriptstyle l = {\rm \,length\,of\,SAP,}} \atop
\scriptstyle q = {\rm \,steps\,from\,} a {\rm \,to\,} b}
\delta_{N {\rm\,mod\,} l, q} W^i \cr
&& + \sum_{{\scriptstyle{\rm SAWs\,}j{\rm \,of\,length\,}N \atop
\scriptstyle {\rm from\,} a {\rm\,to\,} b} \atop
\scriptstyle{\rm no\,}\parallel{\rm \,self-contacts}}
{W^j}.
\label{ensemble}
\end{eqnarray}
In the above each SAP should actually be summed twice, once with
distance $q$ and once with distance $l - q$.  The $\delta$-function in
the SAW part ensures that the particle location after $N$ steps is
$b$.  Sections II and III carry out the disorder average to calculate
the weights $W^{i,j}$ exactly for cases of interest.  The weights turn out
to have a natural interpretation in terms of self-interacting
polymers.

In section II we demonstrate that the trajectories of the lattice
model at the critical energy $E$ are related to a chiral polymer
model, whose exact critical properties are found.  The property of
diffusion at the critical energy is shown to follow from the critical
exponents $\gamma = \frac{6}{7}$ and $\nu_\theta=\frac{4}{7}$ of the
polymer problem.  Then in section III we modify the model so that the
energy of a trajectory is a function of initial position rather than
an independent quantity, and study the localization exponent $\nu$.
An inequality is derived which connects $\nu$ to the exponents of the
associated polymer problem.  Finally, in section IV we discuss
modifications resulting from quantum-mechanical effects, which are
most clearly apparent in scaling laws for higher moments of the
particle distribution function.

\section{Trajectories at the critical energy}

At the critical energy, the particle motion is diffusive:
$\langle\langle R^2(N) \rangle\rangle \sim N$ for long times $N$.
This section shows that the conditional probability $P(a,b,N,E_0)$ for
the particle moves from $a$ to $b$ after $N$ steps,
given that the particle energy is the critical energy $E_0$, is related
to critical properties of self-interacting polymers at the $\theta$-point.
Then in the following section the same mapping will be shown to give
information about trajectories at other energies, and hence about
$\nu$.  We note in passing that in the percolation picture, diffusion at
the critical energy is somewhat surprising.  A particle on the hull of
the infinite cluster moves superdiffusively, while a particle on a
finite cluster has only bounded motion: the diffusive motion obtained
after averaging over initial position essentially
interpolates between these two limits.

For a path $P_{AB}$ at the critical energy $E_0$, the
probability that $P_{AB}$ is the trajectory in a random
potential realization is proportional to $2^{-H_L} 2^{-H_R} =
2^{-H}$.  Here $H_L$ ($H_R$) is the number of hexagons passed
on the left (right) by the path, and $H = H_L + H_R$: the probability
$2^{-H}$ comes about because each hexagon $i$ with potential $V_i$
is as likely to have $V_i > E_0$ as to have $V_i < E_0$.
Then the ensemble (\ref{ensemble}) becomes
\begin{eqnarray}
P(a,b,N,E_0) &=& \sum_{{\scriptstyle{\rm SAPs\,through\,}
a {\rm\,and\,} b,\atop
\scriptstyle l = {\rm \,length\,of\,SAP,}} \atop
\scriptstyle q = {\rm \,steps\,from\,} a {\rm \,to\,} b}
{\delta_{N {\rm\,mod\,} l, q} \over 2^H} \cr
&& + \sum_{{\scriptstyle{\rm SAWs\,of\,length\,}N \atop
\scriptstyle {\rm from\,} a {\rm\,to\,} b} \atop
\scriptstyle{\rm no\,}\parallel{\rm \,self-contacts}}
{1 \over 2^H}.
\end{eqnarray}

The connection to self-interacting polymers appears because the number
of hexagons visited by an SAW is related to the number of
self-contacts of the SAW.  A self-contact is a point where the SAW is
within one edge of intersecting itself.  Counting hexagons in lieu of
self-contacts gives rise to the famous
$\theta^\prime$ model~\cite{coniglio,duplantier} of a two-dimensional
self-interacting polymer.  The number of hexagons visited by an SAW of
length $N$ is $H = N+1-N_2-2N_3$, where $N_2$ and $N_3$ are the
numbers of hexagons visited twice and thrice by the SAW.  Checking
possible paths on the lattice shows that $H = N+1-I-I^\prime$, where
$I$ is the number of self-contacts and $I^\prime$ the number of a
certain type of next-nearest-neighbor contacts.  The effects of
$I^\prime$ are not believed to alter the universality class of the
model~\cite{coniglio,duplantier} and will be ignored.  The weight
$2^{-H} = 2^{-N-1+I}$ thus corresponds to the grand-canonical ensemble
for polymers at chemical potential $\mu = -\log 2$ and with an
attractive interaction energy $\beta U = -\log 2$ for each
self-contact.  We call a self-contact parallel (antiparallel) if, once
a direction is defined along the polymer, the two sections of polymer
in contact have the same (opposite) direction.  For a long polymer,
almost all self-contacts are antiparallel, as might be expected since
parallel self-contacts are a boundary effect, in the weak sense that a
closed polymer has none.

There are three phases of the $\theta^\prime$ model for a
two-dimensional self-interacting polymer.  At high temperature, the
statistical properties are those of the noninteracting SAW, and the
mean radius of gyration is $R \sim N^{3/4}$.  At low temperature,
the polymer is in a collapsed phase with $R \sim N^{1/2}$.  There
is a tricritical point, called the $\theta$-point, separating these
two behaviors, with $R \sim N^{4/7}$.  The importance of
the $\theta$-point for the plateau transition is that
the weight $2^{-H}$ corresponds exactly to the $\theta$-point
on a honeycomb lattice.  The chirality constraint will be shown
to change the scaling and give the same universal properties as
the solvable Manhattan lattice $\theta$-point.

Now the diffusion at the critical energy can be obtained from (\ref{ensemble}).
Considering for the moment only the SAW term in (\ref{ensemble}), the
mean particle displacement after $N$ steps is
\begin{eqnarray}
\langle\langle R^2(N) \rangle\rangle
&=& \sum_b P(a,b,N) ({\bf x}_b - {\bf x_a})^2 \cr
&=&
\sum_{{\rm SAWs\,of\,length\,}N} {R_{\rm SAW}^2 \over 2^H}
\sim \mu^N N^{\gamma-1+2 \nu_\theta}.
\label{dist}
\end{eqnarray}
Here we have introduced the standard polymer exponents $\gamma$ and
$\nu_\theta$, defined through
\begin{eqnarray}
\sum_{{\rm SAWs\,of\,length\,}N} {1 \over 2^H} &\sim& \mu^N N^{\gamma-1}\cr
{\sum_{{\rm SAWs\,of\,length\,}N} {R_{\rm SAW}^2 \over 2^H}
\over \sum_{{\rm SAWs\,of\,length\,}N}{1 \over 2^H}} &\sim& N^{2 \nu_\theta}.
\end{eqnarray}
For ordinary polymers (no chirality constraint) at $\theta$, $\mu =
1$, $\gamma = \frac{8}{7}$, and $\nu_\theta = \frac{4}{7}.$ The effect
of the chirality constraint is clearly to reduce $\gamma$, since some
polymers are forbidden: in fact we now show that $\gamma =
\frac{6}{7}$ with the chirality constraint ($\mu$ and $\nu_\theta$ are
unchanged), so that $\langle\langle R^2(N) \rangle\rangle \propto N$
in (\ref{dist}) and motion is diffusive at the critical energy.

The critical properties of chiral polymers at $\theta$ are actually
related in a very simple way to those of ordinary polymers at
$\theta$.  As shown at the end of this section, the transfer matrix
for $L$ chiral polymers on a cylinder of finite circumference $N$
hexagons has the same leading eigenvalue as the transfer matrix of
$2L$ nonchiral polymers on the same cylinder.  This means that the
critical exponents for the chiral model can be deduced from the known
values for the nonchiral model.

The ``watermelon'' exponents $x_L$~\cite{nienhuis,duplantier} are
defined from the correlation functions $G_{n,L}(a,b)$ of $L$
mutually avoiding SAW's from $a$ to $b$ at criticality: $G_{n,L}(a,b) \propto
|a-b|^{-2 x_L(n)}.$  The natural generalization for the chiral case
is that the $L$ self-avoiding walks have no parallel self-contacts.
Then the computation below of the transfer matrix at criticality ($\mu = 1$)
on finite strips shows that the chiral exponents ${\tilde x}_L$ are
identical to the nonchiral exponents $x_{2L}$ for twice as many
connectors.  The nonchiral values $x_{L} = (L^2 - 1)/12$
known from the Coulomb-gas technique~\cite{nienhuis} then determine all
the ${\tilde x}_L$.

Now we connect the watermelon exponents to physical
properties such as $\gamma$ and $\nu_\theta$.  First,
the size exponent $\nu_\theta = (2 - x_2)^{-1}$ is
unchanged by the chirality constraint because it takes the same value
for ring polymers as for linear polymers, and ring polymers are
unaffected by the chirality constraint.  The exponent $\gamma$ is given
by $\nu_\theta (d - 2 {\tilde x}_1) = \nu_\theta (d - 2 x_2) = \frac{6}{7}.$
Note that the ring exponent $\alpha$
is~\cite{duplantier}
also $\frac{6}{7}$ so the two terms of (\ref{ensemble}) scale with the same
power of $N$, as required for consistency.

The diffusion result
$\langle\langle R^2(N) \rangle\rangle \propto N$ might seem almost
coincidental.  However, it follows directly from the relationship
${\tilde x}_1 = x_2$ between the chiral exponent with one leg and
the nonchiral exponent with two legs:
\begin{equation}
\gamma + 2 \nu_\theta - 1 =
{4 - 2 x_2 \over 2 - x_2} - 1 = 1. 
\end{equation}
The ``mysterious cancellation of exponents''~\cite{gurarie} which yields
diffusion in the percolation picture is relatively simple in the
polymer picture, and does not depend on the specific value of $x_2$.

\begin{figure}
\epsfxsize=2.0truein
\centerline{\epsffile{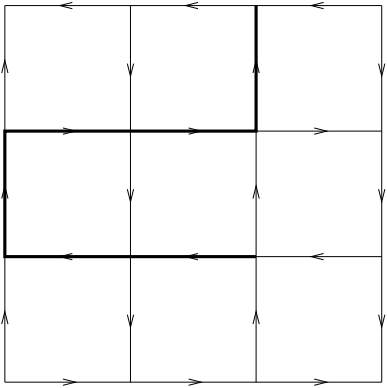}}
\vspace{3mm}
\caption{A directed walk on the Manhattan lattice.  Note that
all allowed self-contacts on this lattice are antiparallel.}
\label{figthree}
\end{figure}

The result ${\tilde x}_L = x_{2L}$ for chiral polymers is exactly the
same as for polymers at $\theta$ on the Manhattan directed lattice
(Fig.~\ref{figthree}), which by construction has no parallel
self-contacts~\cite{bradley}.  Hence we learn that the detailed
structure of the Manhattan lattice is in some sense irrelevant: it is
the short-ranged constraint of no parallel self-contacts which
determines the universality class.  Another piece of information about
polymers follows from the beautiful result of Cardy~\cite{cardy} for
the conductivity $\sigma_{xx} = \frac{\sqrt{3} e^2}{4 h}$ at the
critical energy.  This fixes the lattice diffusion constant through
the Einstein relation~\cite{evers,gurarie}, and therefore predicts a
value for the combination of prefactors in (\ref{dist}).

The chiral polymer model discussed here is just one point of a
two-parameter family of models with antiparallel self-contacts
weighted by some real number $w$ and parallel self-contacts by some
possibly different number $v$.  Then $w=v$ gives ordinary
two-dimensional self-interacting polymers, while $v=0$ gives the
chiral polymer ensemble.  One expects a ``coiled'' polymer phase for
$w=0$ and $v \rightarrow \infty$, different from the conventional
collapsed polymer phase.  The full phase diagram of these models in
the $(w,v)$ plane is a rich subject; the corresponding problem defined
in terms of parallel and antiparallel self-contacts, rather than
hexagons, has been investigated numerically on the square
lattice~\cite{trovato,prellberg2}.  A conformal field theory approach
suggests the possibility of continuously varying $\gamma$ between the
chiral polymer and ordinary polymer $\theta$-points~\cite{cardy2}.  We
remark in passing that the lattice $\theta^\prime$ model defined here
in terms of hexagon weights $(w,z)$ has a number of advantages for
this problem: the critical point $w=\frac{1}{2}$ is known exactly, and
the exact relation discussed below between transfer matrices suggests
that this model may be solvable by vertex methods.

The remainder of this section establishes the connection ${\tilde x}_L
= x_{2L}$ between the watermelon exponents of chiral polymers on the
hexagonal lattice and those of nonchiral polymers, and can be skipped
by nonspecialists.  A powerful method to obtain properties of polymer
models is by using conformal invariance to analyze the results of
finite-size transfer matrix calculations~\cite{derrida,saleur}.  Since
there are reviews of the technique~\cite{saleur,vanderzande}, some
minor details will be omitted.

The goal will be to find the transfer matrix for $L$ polymers on a
cylinder of circumference $h$ hexagons.  The scaling dimensions
${\tilde x}_L$
can be derived from the finite-size correlation length $\xi_{L,h}$,
which is determined by the largest eigenvalue $\lambda_{L,h}$ of the
transfer matrix:
\begin{equation}
\xi_{L,h} = -{1 \over \log \lambda_{L,h}}.
\end{equation}
The ${\tilde x}_L$ are then estimated for successively larger cylinders
using
\begin{equation}
{\tilde x}_{L,h} = - \left({2 \over \sqrt{3}} \right) {h \over 2 \pi }
\log \lambda_{L,h}.
\label{eigconn}
\end{equation}
The geometrical factor $\frac{2}{\sqrt{3}}$ comes from the hexagonal
lattice dimensions and is 1 for a square lattice.

The transfer matrix acts on ``configurations'' of horizontal edges.
A configuration consists of the state of all the horizontal edges,
plus information on which oppositely directed edges are paired
(originated from the same loop), plus information on which hexagons
between horizontal edges have been passed on the left or right.
The entries in the matrix sum over the different possible states of
the vertical edges which can link two configurations.  Each entry
is weighted by a factor $w$ for each hexagon passed on one side,
$v$ for each hexagon passed on both sides, and $\mu$ for each edge.
For the critical point of chiral polymers, $w=\frac{1}{2}$, $v=0$,
and $\mu = 1$; setting $v=w=\frac{1}{2}$ gives the ordinary $\theta$-point,
while taking $w=v=1$ and $\mu = \mu_{c} = (2 + \sqrt{2})^{-1}$ gives
the critical point of noninteracting SAWs.

Note that parallel self-contacts can only occur for one polymer in the
cylindrical geometry when the polymer winds around the cylinder.  As a
result the surface critical exponents, which follow from the transfer
matrix on the strip (closed boundary conditions) rather than on the
cylinder (periodic boundary conditions), are unmodified from the
nonchiral case.  Upon conformal mapping from the cylinder back to the
plane, polymers which wrap around the cylinder become polymers which
wrap around the origin of the plane, and closed boundary conditions
correspond to a branch cut which polymers cannot cross. The
equivalence of surface exponents to those of the nonchiral
$\theta$-point was previously obtained for the Manhattan
lattice~\cite{dequeiroz}.

\begin{table}
\begin{center}
\begin{tabular}{|c| ll | ll |}
$h$&\makebox[10mm]{${\tilde x}_{1,h}=x_{2,h}$}&
\makebox[15mm]{$x_{\rm extrap.}$\hfill}&
\makebox[10mm]{${\tilde x}_{2,h}=x_{4,h}$}
&\makebox[15mm]{$x_{\rm extrap.}$\hfill} \\
\cline{1-5}
&&&&\\
2&0.254768&&& \\
3&0.259127&0.23788&& \\
4&0.256212&0.24859&1.52861&1.2814 \\
5&0.254221&0.24911&1.39654&1.26398\\
6&0.253007&&1.34308&1.25603 \\
7&0.25224&&1.31513& \\
8&&&1.2984& \\
&&&&\\
$\infty$&$\frac{1}{4}=0.25$&&$\frac{5}{4}=1.25$ & \\
&&&&\\
\end{tabular}
\end{center}
\caption{Results of transfer matrix calculations for one and two
chiral polymers (identical to results for two and four nonchiral polymers),
on cylinders of circumference $h$ hexagons.  The largest eigenvalue
of the transfer matrix is related to $x_{L,h}$ through (\ref{eigconn}).
The convergence to the
predicted values ${\tilde x}_1 = \frac{1}{4}$ and
${\tilde x}_2 = \frac{5}{4}$ is seen to be quite
rapid.  The extrapolated values
are obtained by using three consecutive values of $x_h$
to fix the constants in $x_h = c_1 + c_2 h^{-c_3}$, then taking
$c_1$ as an estimate of $x_\infty$.}
\label{tab1}
\end{table}

Table I gives the estimated scaling dimensions ${\tilde x}_1$ and
${\tilde x}_2$ from cylinders of various sizes.  The first few
can be done by hand, while the larger matrices are done by computer.
The leading eigenvalues are exactly the same as for those of twice as
many nonchiral polymers.  This connection is in retrospect not too
surprising, since the condition of no parallel self-contacts for a
polymer from $A$ to $B$ in the chiral case is exactly the condition
that another polymer can be added from $A$ to $B$ in the nonchiral
case.  The subleading eigenvalues can differ, however, so there may not
be a simple equivalence between states of $L$ chiral polymers and $2L$
nonchiral polymers.  The critical properties of the nonchiral model
follow from Coulomb gas results for the $O(n)$ model~\cite{nienhuis},
so we have
\begin{equation}
x_{L} = {L^2 - 1 \over 12},\quad {\tilde x}_L = {4 L^2 - 1 \over 12}.
\end{equation}
The chiral exponents ${\tilde x}_L$ are the same
as those of the $\theta$-point on the Manhattan lattice~\cite{prellberg}.
There are Monte Carlo results for another hexagonal lattice model
believed to lie in the Manhattan
universality class, the ``smart kinetic growth
SAW''~\cite{bennett}, which are consistent with the above values.

\section{The classical localization exponent}

When the particle energy $E$ moves away from the critical energy,
the trajectories become less extended and the mean distance from
the origin after $N$ time steps is reduced.  In this section, we
take the lattice disorder average in order to express $P(a,b,N)$, the
probability that after $N$ steps the particle has moved from $a$ to
$b$, as a weighted sum over linear and ring polymers.  Simple
properties of the weight function then yield an inequality connecting
the localization exponent $\nu$ to polymer exponents $\gamma$ and
$\nu_\theta$.

As in the preceding section, we fix an open or closed curve on the
lattice and ask what fraction of potential realizations make this
curve the correct trajectory.  The particle energy $E$
is also varied in order to find the energy-integrated diffusion constant, and
hence $\nu$.  The weight of a curve is determined by the numbers of
hexagons touched by the curve to the left and right: the requirement
for a path to be the correct trajectory is that all the hexagons to
the right lie above the particle energy.
The probability that a path is the particle trajectory is a function
of the number of different hexagons visited by the path.  The
requirement is that all the hexagons to the immediate left have
energies larger than the particle energy, while those to the immediate
right have energies smaller than the particle energy.  The $H_L$ hexagons
on the left must have higher energies than the $H_R$ on the right,
which is true for $({H_L+H_R \atop H_L})^{-1}$ of potentials.  Furthermore,
the particle energy must lie in the window of width $\sim (H_L + H_R)^{-1}$
between the lowest potential on the left and the highest potential on the
right.  So the weight of an allowed path $P_{AB}$ is
\begin{equation}
W(P_{AB}) \propto {1 \over (H_L+H_R) ({H_L+H_R \atop H_L})} \approx
{H_L! H_R! \over (H_L + H_R + 1)!}.
\label{cpw}
\end{equation}
The same result is obtained by integrating the probability
$(\frac{1+E}{2})^{H_L} (\frac{1-E}{2})^{H_R}$ over particle energy $E$
to obtain a beta function.

Now we can again connect the expression (\ref{ensemble}) and the weight
(\ref{cpw}) to known properties of polymers.  For fixed $H = H_L+H_R$,
the weight is minimized if $H_L = H_R$, and using Stirling's approximation
is then $W(H) \approx {\sqrt{2 \pi} \over 2^H H^{1/2}}$.  The probability
to get from $a$ to $b$ after $N$ steps thus satisfies
\begin{eqnarray}
P(a,b,N) &\geq& \sum_{{\scriptstyle{\rm SAPs\,through\,}
a {\rm\,and\,} b,\atop
\scriptstyle l = {\rm \,length\,of\,SAP,}} \atop
\scriptstyle q = {\rm \,steps\,from\,} a {\rm \,to\,} b}
{\delta_{N {\rm\,mod\,} l, q} \over 2^H H^{1/2}} \cr
&& + \sum_{{\scriptstyle{\rm SAWs\,of\,length\,}N \atop
\scriptstyle {\rm from\,} a {\rm\,to\,} b} \atop
\scriptstyle{\rm no\,}\parallel{\rm \,self-contacts}}
{1 \over 2^H H^{1/2}},
\label{lbound}
\end{eqnarray}
up to a possible numerical constant.  The typical number of hexagons
$H$ scales linearly in $N$ to sufficient accuracy that $H^{1/2}$ can
be replaced by $N^{1/2}$ (this is verified numerically by Monte Carlo
simulations, and if false would require an unexpected multifractality
at the $\theta$-point).  Summing over final positions $b$ to find the
mean squared displacement then gives
\begin{equation}
\langle\langle R^2(N) \rangle\rangle \geq N^{\gamma - 1 + 2 \nu_\theta
- \frac{1}{2}}.
\end{equation}
Then from equation (\ref{polynu}) we obtain an inequality
connecting the localization exponent $\nu$ for the plateau transition to
polymer exponents $\nu_\theta$ and $\gamma$:
\begin{equation}
1 - {1 \over 2 \nu} \geq \gamma + 2 \nu_\theta - \frac{3}{2}.
\label{ineq}
\end{equation}

For the chiral polymer model, the resulting prediction is $\nu \geq
1$, which is satisfied by the actual value $\nu = \frac{4}{3}$.  The
usual nonchiral polymer exponents at $\theta$ would predict $\nu \geq
\frac{7}{3}$ (cf. section IV), so again it is seen that the chirality
constraint is essential.  The fact that the lower bound is not reached
shows that even in the limit of long paths, the number of hexagons to
the left and right of the path cannot be assumed equal in calculating
the weight (\ref{cpw}).  At the critical energy (section II), hexagons
to the left and right contribute equally and this difference is
irrelevant, but away from the critical energy the difference affects
the scaling.

The exact value $\nu = \frac{4}{3}$ is derived in the percolation
picture~\cite{trugman} from the equivalence of closed trajectories at
energy $E$ to percolation hulls with $p-p_c \propto E-E_0$, where
$p_c$ is the critical probability for percolation and $E_0$ is the
critical energy.  Such percolation hulls~\cite{dennijs} have average
size $\xi \sim (p-p_c)^{-\frac{4}{3}}$.  We remark in passing that the
value $\nu = \frac{4}{3}$ can be understood in the polymer context
from the fact the crossover exponent of the tricritical $\theta$-point
is $\phi=\frac{3}{7}$ (this value was first obtained using the
connection to percolation~\cite{duplantier}): then $N^{-\phi} \sim
(p-p_c)$ and $\xi \sim (p-p_c)^{-\nu_\theta/\phi} =
(p-p_c)^{-\frac{4}{3}}$.  The reasons for stressing the inequality
(\ref{ineq}) here rather than the exact result are that the inequality
follows immediately from the classical path weight on polymers and can
be used to gain information on higher moments (Section IV).

The main result of this section is that the exact path weight induced
by averaging over disorder and particle energy can be calculated for
the classical lattice model.  This weight yields the inequality
(\ref{ineq}) connecting statistics of chiral polymers to the critical
exponent $\nu$ of the classical plateau transition.  The focus of
the next section will be whether a similar relationship to polymers
exists for the quantum plateau transition.

\section{Higher moments and the quantum transition}

The quantum Hall plateau transition shows several qualitative
similarities with the semiclassical limit studied in the preceding
sections of this paper.  Both the quantum transition and the
semiclassical limit have power-law delocalization at the critical
energy and a finite critical conductivity.  However, the quantum
transition has proved much more difficult to describe theoretically,
and remains a major open problem.  A natural question is whether any
generalization of the polymer mapping developed for the classical
limit would serve as a useful approach for the quantum case.  The goal
of this section is to show that the diffusive behavior usually assumed to
exist up to the localization length in the quantum case would be
inconsistent with almost any such generalization, and present a
numerical method and preliminary results to verify this assumption.
We focus on one generalization in particular (to {\it nonchiral}
polymers at $\theta$, for reasons described below) for conciseness.

The scaling laws of moments of the particle distribution function
\begin{equation}
\langle R^{2n}(t)\rangle \sim t^{\alpha(n)}
\label{momentlaw}
\end{equation}
demonstrate an essential difference between the classical limit and
the conventional picture of the quantum case.  In this section
$\langle \rangle$ indicates averaging over particle energy and random
potentials, while $\langle \rangle_E$ indicates averaging over random
potentials at fixed particle energy $E$.  If particle motion is
essentially diffusive on short length scales in the quantum case, then
higher scaling laws beyond $n=1$ in (\ref{momentlaw}) do not contain
additional information.  As discussed in the previous section, the
mean squared displacement $\langle R^{2n}(t)\rangle \sim t^{1 -
\frac{1}{2 \nu}}$ contains the localization exponent $\nu$.
This formula was obtained from the assumption that $\langle R^2(t)
\rangle_E$ increases linearly in time at each energy until the
localization length is reached ($R^2 \approx \xi(E)^2$), then
saturates.

If the particle motion is truly diffusive up to the localization
length, then $\langle R^{2n}(t) \rangle_E \sim t^n$ until the
localization length is reached, and
\begin{equation}
\langle R^{2n}(t)\rangle \sim t^{n - \frac{1}{2 \nu}},
\end{equation}
with the localization length exponent $\nu \approx 2.35 \pm
0.05$~\cite{huckestein}.  So if motion in the quantum case is
diffusive up to the localization length, there are no nontrivial
exponents to be found in higher moments of the particle displacement.

Higher moments in the {\it classical} case show nontrivial scaling, and
consequently highly extended trajectories are much more common in the
classical case than the quantum case, even though the localization
length diverges more rapidly near $E_c$ for the quantum
case.  For walks at the critical energy, it follows
from the results of Section II that
\begin{equation}
\langle R^{2n}(t) \rangle_{E_c} \sim t^{2n \nu_\theta + \gamma - 1} =
t^{(8n - 1)/7}.
\end{equation}
Hence, although the mean square displacement does increase linearly
with time, higher moments have nontrivial power laws because the
particle trajectories are not random walks but instead have ``memory,''
as required for the absence of self-intersections.  The higher moments
are more extended than they would be for simple diffusive motion (random
walks).

A similar result holds for the classical case even when the average is
extended to include the particle energy.  The inequality (\ref{ineq})
derived in Section III between polymer exponents at the chiral
$\theta$-point and the scaling of the second moment $\alpha(1) = {1 -
\frac{1}{2 \nu}}$ from (\ref{momentlaw}):
\begin{equation}
1-\frac{1}{2 \nu} \geq 2 \nu_\theta + \gamma -
\frac{3}{2} = \frac{1}{2}.
\end{equation}
For the classical transition with $\nu = \frac{4}{3}$,
the left side is $\frac{5}{8}$ and the inequality is satisfied.  Similarly
for higher moments $\langle R^{2n}(t)\rangle \sim t^{\alpha_c(n)}$
\begin{equation}
\alpha_c(n) \geq 2 n \nu_\theta + \gamma -
\frac{3}{2} = \frac{1}{2} + \frac{8 (n-1)}{7}.
\end{equation}
Hence for sufficiently large $n$ the classical scaling exponents
$\alpha_c(n)$ are larger than the quantum exponents $\alpha(n) = n -
\frac{1}{2 \nu}$, if the motion in the quantum case is indeed
diffusive.

In the remainder of this section, we consider the question of how
quantum interference keeps the quantum case from being related to a
polymer ensemble in the same way as the classical case.  One polymer
ensemble in particular is attractive for the quantum case because the
ensemble is similar to the classical one and the value $\nu =
\frac{7}{3}$ appears in this ensemble, but this connection predicts
nondiffusive motion up to the localization length.  Monte Carlo
numerics are used to test the assumption of diffusive motion in the
quantum case; if verified this assumption would rule out a simple
connection to polymers.

The approach of the previous section was to generate a positive weight
on electron paths by averaging over disorder with the electron path
held constant.  In lattice models for the quantum transition, it
should be possible to attribute a positive weight to each path on the
lattice, and then these weights may be connected to some polymer
problem, presumably different from the chiral ensemble discussed above
for the classical limit.  The first statement (that there is an
assignment of weights) is somewhat trivial from a mathematical point
of view: there are many paths on the lattice between any pair of
points, and hence given any positive probabilities to reach different
points on the lattice after $N$ steps, there is some assignment of
positive path weights which results in the given probabilities.  The
difficult question is whether there is an assignment of weights which
is physically meaningful and related to some local two-dimensional
theory, as in the classical case.  The next paragraphs define a real
(not necessarily positive) path weight; this weight is positive in the
absence of interference, and the leading interference contribution to
this path weight from ``cooperons'' vanishes, though higher
contributions do not.

For fixed disorder, different paths $W_{AB}^i$ from $A$ to $B$
contribute to the amplitude, and the probability $P_{AB}$ to get from
$A$ to $B$ includes both diagonal terms $|W_{AB}^i|^2$ and cross
terms.  Here and in the following we assume a discretized model for
the quantum case, similar to the lattice model introduced previously
for the classical limit.  We start by considering two paths which do
not cross: then the disorder average generates a random phase which
cancels the cross terms, leaving only the (positive) diagonal terms.
The remaining question is what occurs for intersecting paths; this is
known to be the case of interest for the quantum effects causing weak
localization.

If the cross terms did vanish, then in the discretized model where
$\lambda_D$ is effectively zero, $P_{AB}$ would be a sum over (not
necessarily self-avoiding) paths with some positive weight, the
``quantum path weight'' (QPW).  A real weight can be defined even if
cross terms are present by adding to the direct contribution of each
path half of all its disorder-averaged cross terms with other paths.
The QPW picture can break down if, for energies near the critical
energy, the cross terms become large enough to drive the weight
negative for long paths.  However, the leading interference
corrections vanish upon disorder averaging, and a finite strength of
interference is required to drive the path weight negative, so it is
seemingly possible that a positive QPW exists for the paths near the
critical energy which determine $\nu$.  This motivates the conjecture,
tested in the remainder of this section, that the universal
large-length-scale properties of the plateau transition may be related
to those of some classical generalized polymer model (i.e., a sum over
paths with positive weights), in similar fashion to the relationships
found in sections II and III between the classical percolation limit
and the chiral polymer model.

The remaining step is to determine whether any universality class of
classical polymers can reproduce the weights which follow from
disorder averaging in the quantum case.  It seems worthwhile to
identify possibilities, since exact results have been obtained for
many two-dimensional polymer models by Coulomb gas and CFT techniques.
The QPW should give nearly the classical weight (\ref{cpw}) to paths
which are classically allowed or include a small number of quantum
tunneling events, but should not allow of order $N$ tunneling events
for an $N$-step path since then the motion is simply diffusive even
away from the critical energy.  A speculative possibility for the
quantum transition is the {\it nonchiral} polymer ensemble at
$\theta$.  The expectation that quantum mechanics should allow some
unfavorable steps (but fewer than $\sim N$) matches the fact that a
typical polymer in the nonchiral ensemble has some parallel
self-contacts, but fewer than of order $N$.  A simple argument that
the number of parallel self-contacts is subextensive ($<N$) is that a
ring polymer has no parallel self-contacts, so that parallel
self-contacts are in some sense a boundary property.

A surprise is that the value $\nu = \frac{7}{3}$ (which has attracted
attention as the simplest rational consistent with numerics) appears
from exponents of the nonchiral ensemble.  The inequality (\ref{ineq})
connecting the localization exponent $\nu$ to nonchiral polymer
exponents predicts, since $\gamma = \frac{8}{7}$,
\begin{equation}
1 - {1 \over 2 \nu} \geq \gamma + 2 \nu_\theta - \frac{3}{2} = \frac{11}{14},
\end{equation}
or $\nu \geq \frac{7}{3}$.  Hence the value $\frac{7}{3}$ appears in
the critical properties of a polymer ensemble closely related to the
polymer ensemble describing the classical plateau transition.  The the
lower bound is realized if paths have asymptotically the same number
of hexagons to the left as to the right ($H_R \sim H_L$), which should
be a better approximation for the less convoluted paths in the quantum
case.  As seen below, however, this inequality predicts scaling laws
for higher moments which appear to be ruled out numerically in the
quantum case.  Hence quantum interference seems to be relevant at the
quantum transition even beyond the level of changing path weights.

We note in passing that the desired property of diffusion at the
critical energy does not have any simple interpretation as a statement
about the polymer ensemble, since it is only after integrating over
particle energy that the polymer weights may appear.  This situation
is familiar from the Liouvillian approach to the
transition~\cite{sinova,moore}, where the transition is mapped onto a
different problem which contains the exponent $\nu$ but not the
critical conductivity.  Note that the previous appearance of the value
$\frac{7}{3}$ in a semiclassical average over a single percolation
trajectory~\cite{milnikov} does not clearly relate to a critical
point, when the electron is delocalized over multiple trajectories.
Now we discuss how numerics can test whether the appearance of this
value in this polymer ensemble is just a numerical coincidence.

There are numerically testable consequences which can be used to check
whether nonchiral polymers at $\theta$ are indeed related to the
quantum case.  The polymer problem predicts various moments of the
disorder- and energy-averaged displacement: it was shown in the
previous paragraph that the mean squared displacement $\langle R^2(N)
\rangle \sim N^{11/14}$, so that $\nu = \frac{7}{3}$.  Similar
predictions follow for higher moments, such as $\langle R^4(N) \rangle
\sim N^{4 \nu + \gamma - 3/2} = N^{27/14}$, or $\alpha(2) = \frac{27}{14}$.
This can be compared to the null hypothesis of diffusion up to the
localization length, which predicts $\langle R^4(N) \rangle \sim
N^{1.78}$.

We have performed Monte Carlo simulations with up to 1800 states in
the lowest Landau level to track the evolution of a localized wave
packet in a disordered potential (the method is similar to that
of~\cite{boldyrev}).  The error bars are larger for the fourth moment
than for the second because finite-size effects are more pronounced on
the extended paths which dominate the fourth moment, but it appears
that $\alpha(2) = 1.8 \pm 0.1$, which if correct is sufficient to rule
out $\alpha(2) = 27/14 \approx 1.92.$ With larger system sizes, it
should be straightforward to confirm the assumption of diffusive
motion up to the localization length.  Then a polymer description would
have to have $\nu = 1/2$ (either dense polymers or random walks), but
no appropriate ensemble is obvious.  It seems more likely that strong
quantum interference prevents a physically meaningful
assignment of path weights in the quantum case. 

To summarize, this section discussed differences between the classical
and quantum transitions which become apparent in higher moments of the
particle distribution function.  Numerics seem to support the picture
of diffusion up to the localization length and rule out the simplest
polymer model for the quantum case.  In closing, we mention briefly
connections between the polymer models discussed in this paper and
conformal field theory (CFT) approaches to the transition.  The
low-temperature $O(n)$ phase also appears in a large-$N$ expansion of
the disorder-averaged Liouvillian theory, which is similar to a
partially supersymmetric complex $O(N)$ model, with $N \rightarrow 1$
the physical limit~\cite{moore}.  Recent work~\cite{leclair,bhaseen2}
on the critical point of Dirac fermions in a nonabelian random vector
potential found a $c=-2$ dense polymer problem hidden in the critical
theory for several different types of disorder.  Since upon adding
additional disorder (random mass and chemical potential) the abelian
version of this critical point flows to the plateau transition fixed
point~\cite{ludwig}, the appearance of polymer subalgebras may be
generic to this class of random critical points.

The author wishes to thank H. Saleur and S. Girvin for helpful
suggestions.

\end{document}